\documentclass[showpacs,preprint,aps]{revtex4}%
\usepackage{amsfonts}
\usepackage{amsmath}
\usepackage{amssymb}
\usepackage{epsfig}
\usepackage{graphicx}%
\ifx\pdfoutput\relax\let\pdfoutput=\undefined\fi
\newcount\msipdfoutput
\ifx\pdfoutput\undefined\else
\ifcase\pdfoutput\else
\ifx\paperwidth\undefined\else
\ifdim\paperheight=0pt\relax\else\pdfpageheight\paperheight\fi
\ifdim\paperwidth=0pt\relax\else\pdfpagewidth\paperwidth\fi
\fi\fi\fi
\begin{document}
\title{Stability of Pairwise Entanglement in a Decoherent Environment }
\author{Jian-Ming Cai}
\email{jmcai@mail.ustc.edu.cn}
\author{Zheng-Wei Zhou}
\email{zwzhou@ustc.edu.cn}
\author{Guang-Can Guo}
\email{gcguo@ustc.edu.cn}
\affiliation{Key Laboratory of Quantum Information, University of Science and Technology of
China (CAS), Hefei 230026, People's Republic of China.}

\begin{abstract}
Consider the dynamics of a two-qubit entangled system in the decoherence
environment, we investigate the stability of pairwise entanglement under
decoherence. We find that for different decoherence models, there exist some
special class of entangled states of which the pairwise entanglement is the
most stable. The lifetime of the entanglement in these states is larger than
other states with the same initial entanglement. In addition, we also
investigate the dynamics of pairwise entanglement in the ground state of spin
models such as Heisenberg and XXY models.

\end{abstract}

\pacs{03.67.Pp, 03.65.Ud, 03.65.Yz, 03.67.Mn}
\maketitle

\section{Introduction}

Quantum entanglement is nonlocal and excess-classical correlation between
separate parties, which is a most important character of quantum mechanics
\cite{EPR&Schrodinger&Bell}. Lots of interest has focused on the nature of
entanglement and the structure of entangled states \cite{Nielsen & Chuang}.
Besides, entanglement is also the most central and indispensable resource in
quantum information processing such as quantum computation \cite{Nielsen &
Chuang}, quantum teleportation \cite{Qtele}, quantum dense coding \cite{Qds},
and quantum key distribution \cite{QKD}. In a word, quantum entanglement is
not only importance in theory but also in practical applications.

One the other hand, it is well known that decoherence \cite{Decoherence} is a
vital factor that should not be neglected in quantum information processing.
The coupling between any quantum system and its environment is inevitable.
Thus the entanglement will evidently be reduced and even disappear because of
this system-environment coupling, e.g. in a large scale quantum computer or
during the course of entanglement distribution via noisy channels. The
stability of entanglement depends on the initial entangled system -- its
entanglement structure and its size
\cite{Simon&Kempe,Dur&Briegel,Carvalho&Mintert&Buchleitner,Lidar1}. We may
look on entanglement as a bond between different qubits, just like the
chemical bond between different atoms. People studied the behavior of chemical
bonds in different environment to understand how chemical bonds are formed.
Thus investigating the entanglement dynamics of different types of entangled
states in the decoherence environment may help to gain some insight into the
properties of the decoherence and the entanglement, which will provide useful
hints for maintaining entanglement. And what kind of entanglement bond is the
most stable under different decoherence models is an interesting problem.

In this paper, we investigate the evolution of pairwise entanglement for
two-qubit entangled states in the decoherence model which is described by
general Pauli channels. We use the concurrence of Wootters\cite{ConC}, which
is related straightforwardly to the entanglement of formation (EOF), as the
measure of entanglement for two-qubit entangled states. The most interesting
problem is that given some general decoherence model, what kind of entangled
states can maintain entanglement best. We find that with the same initial
entanglement, the lifetime of entanglement in some specific class of entangled
states is the longest. For a special decoherence model, that is depolarizing
channels, all pure states together with some mixed states, which we call as
Decoherence Path States (DPS) is the most entanglement-stable. We present the
analytic dynamics of two-qubit entanglement for these special entangled
states. Furthermore, we also study the stability of the nearest neighbor
entanglement in the ground state of some spin models such as Heisenberg and
XXY model. Coincidentally, the conclusion is that in some noise models, the
nearest neighbor entanglement in the ground state is also the most stable,
though it is not maximized \cite{Entangled rings}.

The paper is organized as follows. In Sec. II we introduce the entanglement
measure of two-qubit entanglement and the decoherence model, which can be
viewed as a completely positive map. In Sec. III we investigate the dynamics
of two-qubit entanglement under the influence of decoherence and try to find
the special entangled states of which the pairwise entanglement is the most
stable. In Sec. VI we examined the evolution of entanglement for some specific
and maybe important mixed states, e.g. the ground states of spin models and
the maximally entangled mixed states etc. In Sec. IV conclusions and
discussions, together with some interesting open questions are presented.

\section{Entanglement Measure and Decoherence model}

There have been a number of measures for two-qubit entanglement, such as the
entanglement of formation \cite{EOF,EntanglementMeasure}, negativity
\cite{negativity} and relative entropy of entanglement \cite{REE} etc. In this
paper, we adopt the well-established measure of entanglement concurrence as
the measure of two-qubit entanglement. Consider a general two-qubit state, the
density matrix is $\rho$. Then its time-reversed matrix is defined as
\begin{equation}
\widetilde{\rho}=(\sigma_{y}\otimes\sigma_{y})\rho^{\ast}(\sigma_{y}%
\otimes\sigma_{y})
\end{equation}
The concurrence of $\rho$ is given by \cite{ConC}
\begin{equation}
C=\max\left\{  0,\sqrt{\lambda_{1}}-\sqrt{\lambda_{2}}-\sqrt{\lambda_{3}%
}-\sqrt{\lambda_{4}}\right\}
\end{equation}
where $\lambda_{i}$s are the eigenvalues of $\rho\widetilde{\rho}$ in
decreasing order. The corresponding entanglement of formation can be evaluated
as
\begin{equation}
\xi\left(  C\right)  =h\left(  \frac{1+\sqrt{1-C^{2}}}{2}\right)
\end{equation}
where $h\left(  x\right)  =-x\log_{2} x-(1-x)\log_{2}(1-x)$ is the Shannon's
entropy function.

However, it is not a very simple task to calculate the concurrence of a
two-qubit system in an analytic way. Here we adopt a new method of calculating
the entanglement of formation and thus the concurrence, which is based on
Lorentz singular-value decomposition \cite{Verstraete}. For an arbitrary
$2\times2$ state $\rho$, there exists a $4\times4$ matrix with elements
$R_{ij}=Tr(\rho\sigma_{i}\otimes\sigma_{j})$. In the real $R$-picture, the
density matrix $\rho$ can be written as
\begin{equation}
\rho=\frac{1}{4}\sum\limits_{ij=0}^{3}R_{ij}\sigma_{i}\otimes\sigma_{j}%
\end{equation}
where $\{\sigma_{i}\}$ are the Pauli matrices.

\bigskip\emph{Lemma 1:} The $4\times4$ matrix $R$ can be decomposed as
$R=L_{1}\Sigma L_{2}^{T}$, with $L_{1}$, $L_{2}$ finite proper orthochronous
Lorentz transformations given by $L_{1}=T(A \otimes A^{\ast})T^{\dagger}$,
$L_{2}=T(B \otimes B^{\ast})T^{\dagger}$, where $T=\frac{1}{\sqrt{2}}\left(
\begin{array}
[c]{cccc}%
1 & \cdot & \cdot & 1\\
\cdot & 1 & 1 & \cdot\\
\cdot & i & -i & \cdot\\
1 & \cdot & \cdot & -1
\end{array}
\right)  $. The normal form $\Sigma$ is either of real diagonal form
$\Sigma=diag[s_{0},s_{1},s_{2},s_{3}]$ with $s_{0}\geq s_{1}\geq s_{2}%
\geq|s_{3}|$, or of the form
\begin{equation}
\left(
\begin{array}
[c]{cccc}%
a & \cdot & \cdot & b\\
\cdot & d & \cdot & \cdot\\
\cdot & \cdot & d & \cdot\\
c & \cdot & \cdot & b+c-a
\end{array}
\right)
\end{equation}
with $a$, $b$, $c$, $d$ real. And the Lorentz singular values of the second
normal form are given by $[s_{0},s_{1},s_{2},s_{3}]=[\sqrt{(a-b)(a-c)},
\sqrt{(a-b)(a-c)}, d, -d]$.

\bigskip\emph{Lemma 2:} Given a state $\rho$ and the corresponding matrix $R$,
the concurrence of $\rho$ is $C=\max{\{ 0,(-s_{0}+s_{1}+s_{2}-s_{3})/2\}}$
depending on the Lorentz singular values of $R$. And $s_{0}-s_{1}-s_{2}%
+s_{3}=\min\limits_{L_{1}, L{2}}Tr(L_{1}RL_{2}^{T})$, where $L_{1}$, $L_{2}$
are proper orthochronous Lorentz transformations.

Based on the above two useful lemmas, we can see that the concurrence of a
given density matrix $\rho$ is only determined by the Lorentz singular values
of the corresponding $R$-matrix, which are the only invariants of a state
under determinant 1 SLOCC operations \cite{Verstraete}. In the following, we
adopt this method for calculating the residual entanglement of an initial
entangled state in the decoherence environment. It is shown that the influence
of the environment on the pairwise entanglement is reflected by changing the
Lorentz singular values. Before proceeding to the details, we first introduce
the decoherence model generated by Pauli operators.

For a general decoherence model, it can be denoted as a completely positive
map with an operator-sum representation. The effect of the general Pauli
channels on a qubit $\rho$ is described as follows \cite{Nielsen & Chuang}
\begin{equation}
\varepsilon( \rho_{i}) =p_{0}\rho+\sum\limits_{i=1}^{3} p_{i}\sigma_{i}%
\rho\sigma_{i}%
\end{equation}
where $p_{i}\geq0$, $\sum\limits_{i=0}^{3}p_{i}=1$, and $\sigma_{i}$ are Pauli
operators. This decoherence model includes some representative noise channels.
When $p_{1}=p_{2}=p_{3}$ it is just the depolarizing channel, which describes
the decoherence process related to the couplings of quantum system to the
thermal reservoir in the large temperature limit \cite{Dur&Briegel}. And when
$p_{1}=p_{2}=0$ the noise model is the dephasing channel, without energy
exchange between the system and the environment, and only lose phase
information. These kinds of decoherence models are common in several physical
systems. In the rest of this paper, we will investigate how the decoherence
will influence the pairwise entanglement in details.

\section{Dynamics of pairwise entanglement under decoherence}

\subsection{General Pauli Channels}

We fist assume that each qubit is independently coupled to the environment.
The environment is characterized by the noisy channels generated by Pauli
operators as shown in Eq (6). The initial state $\rho$, associated with this
state $R=L_{1}\Sigma L_{2}^{T}$, is an entangled two-qubit state. Then after
some time, $\rho$ will be transformed to another state $\rho^{\prime}$ with
much less entanglement, due to the action of the noisy channels, that is
$\rho^{\prime}=\varepsilon_{1}\varepsilon_{2}(\rho)$. According to Eq (6), it
can be obtained easily that
\begin{equation}
\rho^{\prime}=\sum\limits_{i,j=0}^{3}(M_{i}\otimes N_{j})\rho(M_{i}\otimes
N_{j})^{\dagger}%
\end{equation}
where $M_{i}=\sqrt{p_{i}}\sigma_{i}$ and $N_{j}=\sqrt{p_{j}}\sigma_{j}$. Now
we can transformed this map into the $R$-picture. Denote the $R$-matrix
associated with the state $\rho^{\prime}$ as $R^{\prime}$, then
\begin{equation}
R^{\prime}=\sum\limits_{i,j=0}^{3}L_{M_{i}} R L^{T}_{N_{j}} =(\sum
\limits_{i=0}^{3}L_{M_{i}}) R (\sum\limits_{j=0}^{3}L^{T}_{N_{j}})
\end{equation}
where $L_{M_{i}}$ and $L_{N_{j}}$ are Lorentz transformations given by
$L_{M_{i}}=T(M_{i}\otimes M_{i}^{\ast})T^{\dagger}$ and $L_{N_{j}}%
=T(N_{j}\otimes N_{j}^{\ast})T^{\dagger}$.

For simplification, we can introduce $L_{1}=\sum\limits_{i=0}^{3}L_{M_{i}}$
and $L_{2}=\sum\limits_{j=0}^{3}L_{N_{j}}$. Therefore, the state evolution
under decoherence is simply characterized by $R^{^{\prime}}=L_{1} R L^{T}_{2}$
in the real $R$-picture. For the decoherence model we discussed here,
$L_{1}=\sum\limits_{i=0}^{3}T(M_{i}\otimes M_{i}^{\ast})T^{\dagger}$,
$L_{2}=\sum\limits_{j=0}^{3}T(N_{j}\otimes N_{j}^{\ast})T^{\dagger}$. After
simple calculation, it can be seen that $L_{1}=L_{2}=diag {[1,Q_{1}%
,Q_{2},Q_{3}]}$, where $Q_{1}=p_{0}+p_{1}-p_{2}-p_{3}$, $Q_{2}=p_{0}%
-p_{1}+p_{2}-p_{3}$ and $Q_{3}=p_{0}-p_{1}-p_{2}+p_{3}$. It is obvious that
the action of the noisy channels on the entanglement can be viewed as
shrinking the Lorentz singular values by the above three coefficients.

If the initial entangled states are set as pure states, then according to the
Schmidt decomposition theorem \cite{Nielsen & Chuang}, an arbitrary two-qubit
pure state $| \Omega\rangle$ can be expressed as $| \Omega\rangle=\lambda_{1}|
0^{\prime}1^{\prime}\rangle+\lambda_{2}| 1^{\prime}0^{\prime}\rangle$, where
here $\lambda_{1}$ and $\lambda_{2}$ are non-negative real numbers satisfying
$\lambda_{1}^{2}+\lambda_{2}^{2}=1$. That is there always exist local unitary
operations $U$ and $V$ , which satisfy $| \Omega\rangle=(U \otimes V)|
\Omega_{0} \rangle$, where $| \Omega_{0} \rangle=\lambda_{1}| 01
\rangle+\lambda_{2}|10 \rangle$. Here $| 0 \rangle$, $|1 \rangle$ are the
$+1$, $-1$ eigenstates of the Pauli $\sigma_{z}$ matrix. In the $R$-picture,
$| \Omega_{0} \rangle\langle\Omega_{0}|$ corresponds to the matrix:
\begin{equation}
R_{0}=\left(
\begin{array}
[c]{cccc}%
1 & \cdot & \cdot & \lambda_{2}^{2}-\lambda_{1}^{2}\\
\cdot & 2\lambda_{1}\lambda_{2} & \cdot & \cdot\\
\cdot & \cdot & 2\lambda_{1}\lambda_{2} & \cdot\\
\lambda_{1}^{2}-\lambda_{2}^{2} & \cdot & \cdot & -1
\end{array}
\right)
\end{equation}
And the local unitary operations on $\left\vert \Omega_{0} \right\rangle $
correspond to left and right multiplication of $R_{0}$ with orthogonal
matrices, therefore in the $R$-picture, an arbitrary pure state $\left\vert
\Omega\right\rangle \left\langle \Omega\right\vert $ corresponds to the
matrix:
\begin{equation}
R=L_{U} R_{0} L_{V}^{T}%
\end{equation}
where $L_{U}=(%
\begin{array}
[c]{cccc}%
1 & \cdot &  & \\
\cdot & O_{1} &  & \\
&  &  &
\end{array}
)=T(U\otimes U^{\ast})T^{\dagger}$ and $L_{V}=(%
\begin{array}
[c]{cccc}%
1 & \cdot &  & \\
\cdot & O_{2} &  & \\
&  &  &
\end{array}
)=T(V\otimes V^{\ast})T^{\dagger}$, with $O_{1}$ and $O_{2}$ are real
$3\times3$ orthogonal matrices with determinant $1$.

In the following we will investigate how the pairwise entanglement changes
under decoherence in the real $R$-picture and try to find what kind of
entangled states, with the same initial entanglement, can maintain
entanglement best.

We starting by considering the initial entangled state is in the Schmidt
decomposition form $\left\vert \Omega_{0} \right\rangle $. As discussed above,
due to the coupling between the system and environment, $\left\vert \Omega_{0}
\right\rangle $ is transformed into another mixed states $\rho^{\prime}%
_{0}=\varepsilon_{1}\varepsilon_{2}(\left\vert \Omega_{0} \right\rangle
\left\langle \Omega_{0}\right\vert )$ . In the $R$-picture, this action can be
expressed as $R^{\prime}_{0}=L_{1}R_{0}L^{T}_{2}$, that is:
\begin{equation}
R^{\prime}_{0}=\left(
\begin{array}
[c]{cccc}%
1 & \cdot & \cdot & (\lambda_{2}^{2}-\lambda_{1}^{2})Q_{3}\\
\cdot & 2\lambda_{1}\lambda_{2}Q_{1}^{2} & \cdot & \cdot\\
\cdot & \cdot & 2\lambda_{1}\lambda_{2}Q_{2}^{2} & \cdot\\
(\lambda_{1}^{2}-\lambda_{2}^{2})Q_{3} & \cdot & \cdot & -Q_{3}^{2}%
\end{array}
\right)
\end{equation}
Therefore the concurrence of $\rho^{\prime}_{0}$ can be obtained easily
according to lemma 2.
\begin{equation}
C^{\prime}=\max\{0, \frac{C_{0}(Q_{1}^{2}+Q_{2}^{2})+Q_{3}^{2}-1}{2}\}
\end{equation}
where $C_{0}=2\lambda_{1}\lambda_{2}$ is the initial entanglement. Actually,
this result can also be obtained by the conventional way of calculating the
concurrence. Note that
\begin{equation}
\rho^{\prime}_{0}=\left(
\begin{array}
[c]{cccc}%
A & \cdot & \cdot & B\\
\cdot & D & C & \cdot\\
\cdot & C & E & \cdot\\
B & \cdot & \cdot & A
\end{array}
\right)
\end{equation}
where $A=(1-Q_{3}^{2})/4$, $B=(Q_{1}^{2}-Q_{2}^{2})\lambda_{1}\lambda_{2}/2$,
$C=(Q_{1}^{2}+Q_{2}^{2})\lambda_{1}\lambda_{2}/2$, $D=(1+Q_{3}^{2}%
)/4+Q_{3}(\lambda_{1}^{2}-\lambda_{2}^{2})/2$ and $E=(1+Q_{3}^{2}%
)/4-Q_{3}(\lambda_{1}^{2}-\lambda_{2}^{2})/2$. For a state having a density
matrix of the above form, the concurrence is given by $C^{\prime
}=max{\{0,C_{1},C_{2}\}}$, where $C_{1}=2(|B|-\sqrt{DE})$ and $C_{2}=2(C-A)$.
We note $C_{2}$ is always larger than $C_{1}$, thus $C^{\prime}%
=max{\{0,2(C-A)\}}$ which agrees with the above result.

For an arbitrary pure state $|\Omega\rangle$, the corresponding $R$ matrix is
shown in Eq. (10). In the decoherence environment, $|\Omega\rangle$ is
transformed to $\rho^{\prime}$. In the real $R$-picture, the evolution of the
state is described by
\begin{equation}
R^{\prime}=L_{1} R L^{T}_{2}=L_{1}L_{U} R_{0} L_{V}^{T} L^{T}_{2}%
\end{equation}
We then multiply $L_{U}^{T}$ and $L_{V}$ to $R$ from left and right
respectively, and get another $R$-matrix $R^{\prime\prime}=L^{\prime}_{1}
R_{0} L^{\prime}_{2}$, where $L^{\prime}_{1}=L_{U}^{T} L_{1}L_{U}$ and
$L^{\prime}_{2}=L_{V}^{T} L^{T}_{2} L_{V}$. This corresponds to local unitary
operations on the state $\rho^{\prime}$ in the $\rho$-picture, thus the
concurrence of $R^{\prime\prime}$ is identical to the one of $R^{\prime}$. As
we have discussed above, the concurrence of a given $R$ matrix is only
determined by the Lorentz singular values, and the action of the decoherence
on the entanglement can be viewed as shrinking the Lorentz singular values. If
we look on $(s_{0},s_{1},s_{2},s_{3})$ as the components of a vector, then the
influence of decoherence is just shrinking this vector according to the
coefficients $Q_{1}$, $Q_{2}$ and $Q_{3}$. This can be reflected by the
residual entanglement shown in Eq. (12). The action of the noisy channels is
characterized by the three shrinking coefficients. In addition, we note that
$L_{U}, L_{V}\in SO(3)$. Therefore the effects of $L_{U}$ and $L_{V}$ on
$L_{1}$ and $L_{2}$ respectively is changing the shrinking directions. Then if
we order the coefficients $\{n_{1}, n_{2}, n_{3}\}=\{Q_{1}, Q_{2}, Q_{3}\}$
such that $n^{2}_{1}\leq n^{2}_{2}\leq n^{2}_{3}$. Based on the above
analysis, it can easily be seen that the maximum residual entanglement is
\begin{equation}
C^{\prime}_{max} =\max\{0, \frac{C_{0}(n_{1}^{2}+n_{2}^{2})+n_{3}^{2}-1}{2}\}
\end{equation}
where $C_{0}$ is the initial entanglement. The above maximum residual
entanglement can be achieved by appropriate local unitary operations. The
corresponding initial pure states are
\begin{equation}
\wp=\{|\Omega\rangle=(U\otimes U)|\Omega_{0}\rangle\mid L_{U}^{T}L_{1}%
L_{U}=diag[1,\pm n_{1/2}^{2},\pm n_{2/1}^{2},\pm n_{3}^{2}]\}
\end{equation}
This set of pure states present those special states that are the most
entanglement-stable. The minimum residual entanglement and the corresponding
initial pure states can be also derived in a similar way. Given a specific
decoherence model, the relation between $Q_{1}$, $Q_{2}$, and $Q_{3}$ is
known, then the pure states in $\wp$ can be written explicitly. For example,
in the dephasing channels model, the coefficients are $Q_{1}=Q_{2}=p_{0}%
-p_{3}$ and $Q_{3}=1$. Therefore both $|\Omega_{0}\rangle$ and $(\sigma
_{x}\otimes\sigma_{x})|\Omega_{0}\rangle$ belong to the set $\wp$. The other
states contained in $\wp$ can also be obtained easily according Eq. (16).

Although all the pure states with the same entanglement is equivalent under
local unitary operations, it can be seen from the above results that under
decoherence the behavior of different pure states are not all the same. This
reflect the properties of the decoherence model and its influence on the
entanglement. Our results suggest that if the decoherence model is fixed,
there exists a special class of pure states, with the same initial
entanglement, which are more favorable for maintaining entanglement. This
gives some useful hints for maintaining and distributing entanglement. For
example, if we want to distribute an entangled pure state between two separate
parties through noisy channels, it will be helpful to apply some local unitary
operations beforehand to transform the entangled pure state into the form of
the states in the above set $\wp$.

For an initial general mixed entangled state $\rho$ with the initial
entanglement $C_{0}$, to drive an analytic evolution equation for its
entanglement maybe intractable. However, we derive a upper bound of the
residual entanglement in the following. This upper bound is the corresponding
residual entanglement $C_{\wp}^{\prime}$ for the states in the set $\wp$ with
the same initial entanglement $C_{0}$. It has be shown in \cite{ConC}, there
exists an optimal decomposition $\rho=\sum\nolimits_{i}p_{i}|\varphi
_{i}\rangle\langle\varphi_{i}|$, such that $C_{i}=C_{0}$ for each
$|\varphi_{i}\rangle$, with $\sum\nolimits_{i}p_{i}=1$. Then under decoherence
$\rho$ is transformed into another state $\rho^{\prime}=\sum\nolimits_{i}%
p_{i}\varepsilon_{1}\varepsilon_{2}(|\varphi_{i}\rangle\langle\varphi_{i}|)$.
According to the convexity of the concurrence, we know that $C^{\prime}%
\leq\sum\nolimits_{i}p_{i} C^{\prime}_{i}$. Since $C^{\prime}_{i}\leq C_{\wp
}^{\prime}$, it is obvious that $C^{\prime}\leq C_{\wp}^{\prime}$. This
suggests that the state in the set $\wp$ is the most entanglement-stable of
all the states with the same initial entanglement, no matter pure or general
mixed states. It is well known that, an arbitrary pure state can be
transformed into the states, with the same entanglement, of the set $\wp$ by
local unitary operations. Thus in the sense discussed in this paper, the
pairwise entanglement in pure states are more favorable for maintaining
entanglement compared to the generic mixed states. This conclusion is valid
for any decoherence model which can be verified from the above discussions.

Up till now, we have investigate the entanglement dynamics of two-qubit states
assuming that each qubit is independently coupled to the environment. In the
following, we consider the situation that only one qubit is under decoherence
and the initial state is a pure entangled state. We are interested in whether
the lifetime of entanglement also depends on the initial entanglement in this
situation. As discussed above, the entanglement dynamics for pure states with
the same initial entanglement are also dependent on their forms. To be
comparable, we set the initial pure state in the Schmidt decomposition form $|
\Omega_{0} \rangle=\lambda_{1}| 01 \rangle+\lambda_{2}|10 \rangle$. If only
the first qubit is under decoherence, then the two-qubit system becomes
$\rho^{\prime}=\varepsilon_{1}(| \Omega_{0} \rangle\langle\Omega_{0}|)$. In
the $R$-picture, the matrix corresponding to $\rho^{\prime}$ is $R^{\prime
}=L_{1}R_{0}$, that is
\begin{equation}
R^{\prime}=\left(
\begin{array}
[c]{cccc}%
1 & \cdot & \cdot & \lambda_{2}^{2}-\lambda_{1}^{2}\\
\cdot & 2\lambda_{1}\lambda_{2}Q_{1} & \cdot & \cdot\\
\cdot & \cdot & 2\lambda_{1}\lambda_{2}Q_{2} & \cdot\\
(\lambda_{1}^{2}-\lambda_{2}^{2})Q_{3} & \cdot & \cdot & -Q_{3}%
\end{array}
\right)
\end{equation}
Therefore the concurrence of $\rho^{\prime}$ is
\begin{equation}
C^{\prime}=2\max{\{0,\frac{C_{0}}{2}(|Q_{1}-Q_{2}|-Q_{3}-1),\frac{C_{0}}%
{2}(|Q_{1}+Q_{2}|+Q_{3}-1)\}}%
\end{equation}
This result shows that the lifetime of the entanglement is independent on the
initial entanglement. No matter how much the initial two-qubit state is
entangled, it becomes separable in a constant time. This somewhat interesting
phenomena reflect that entanglement is some kind of nonlocal property.

\subsection{Depolarizing Channels}

In the above, we discuss the dynamics of entanglement under the noisy channels
generated by Pauli matrices. When the parameters satisfy $p_{1}=p_{2}%
=p_{3}=p/4$ and $p_{0}=1-3p/4$, the decoherence model in Eq. (6) is the
depolarizing channels. The depolarizing channels describe the
system-environment coupling in the large temperature limit $T\rightarrow
\infty$. It can be realized by random Von Neumann measurements. Again we
assume that each qubit is independently coupled to the environment. The
shrinking coefficients are $Q_{1}=Q_{2}=Q_{3}=1-p$. Taking into account of the
strength of the system-environment coupling and the interaction time, we can
write $1-p(t)=e^{-\kappa t}$ \cite{Dur&Briegel}. Thus the corresponding
matrices in the $R$-picture are $L_{1}=L_{2}=diag[1, e^{-\kappa t}, e^{-\kappa
t}, e^{-\kappa t}]$. Given an arbitrary initial entangled pure state
$|\Omega\rangle$, the corresponding Schmidt decomposition normal form is
$|\Omega_{0}\rangle$ associated with the $R$-matrix $R_{0}$. Then after time
$t$, the residual entanglement is dependent on the $R^{\prime\prime}%
=L_{1}R_{0}L_{2}^{T}$. Note that $L_{U}^{T}L_{1}L_{U}=L_{1}$ and $L_{V}%
^{T}L_{2}L_{V}=L_{2}$ here. Therefore the concurrence at time $t$ is
\begin{equation}
C^{\prime}(t)=\max\{0, (C_{0}+\frac{1}{2})e^{-2\kappa t}-\frac{1}{2}\}
\end{equation}

From the above evolution function of the pairwise entanglement, we can find
that the residual entanglement at time $t$, only depends on its initial
entanglement $C_{0}$. Thus for all two-qubit pure states coupled with the same
depolarizing environment, the stability of the entanglement is only governed
by their initial entanglement, although these pure states could be in
different forms. In other words, all pure states are the most
entanglement-stable, need not to apply local unitary operations beforehand.
Recalling the above analysis in the $R$-picture, the reason for this
interesting result is that the shrinking of the Lorentz singular values of the
associated $R$-matrix, which is introduced by the depolarizing channels, is
isotropic. Besides, it is also obviously that even for two generic mixed
state, if there are $LU$ equivalent then the dynamics of two-qubit
entanglement are also equal. Furthermore, there are some special mixed
entangled states which have the same entanglement dynamics as pure states,
that is also the most entanglement-stable. We will discuss the situation of
mixed states in the next section. In Fig. 1 we present a visual example by
plotting the dynamics of residual entanglement for pure states and some other
generic mixed states with the same initial entanglement.

\begin{figure}[tbh]
\includegraphics[scale=0.5]{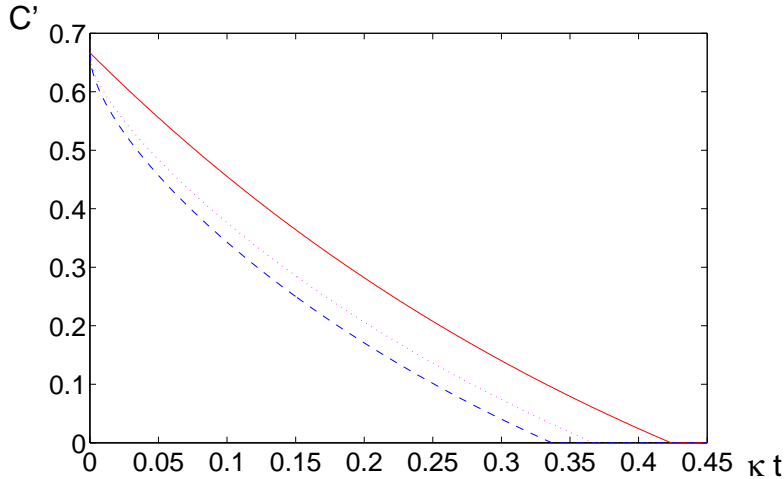}\caption{(Color online) The dynamics
of residual entanglement in a two-qubit system. The initial states are chosen
as pure states and some other generic mixed states with the same initial
entanglement. We set $C_{0}=2/3.$ Residual entanglement $C^{\prime}$ as
function of time $t$. Pure state (Solid Curve); some other generic mixed
states (Dashed and Dotted Curve).}%
\end{figure}

\section{Examples: mixed states}

\subsection{Decoherence Path States}

\bigskip\emph{Definition:} Given a decoherence model characterized by a
completely positive map $\Lambda$, Decoherence Path States (DPS) are those
transient states $\rho$ obtained from the pure states, that is $\exists$
$|\psi\rangle$ which satisfies $\rho=\Lambda(|\psi\rangle\langle\psi|)$.

We consider the depolarizing channels, and the initial entangled state $\rho$
is a decoherence path state with the initial entanglement $C_{0}\geq0$. Thus
there exists $|\psi\rangle$ with the entanglement $C$ and some time $t_{0}$
that satisfy $\rho=\Lambda(t_{0}) (|\psi\rangle\langle\psi|)$ and
$C_{0}=(C+\frac{1}{2})e^{-2\kappa t_{0}}-\frac{1}{2}$. Then after some time
$t$, the decoherence path state $\rho$ evolves to another state $\rho^{\prime
}(t)$ with the entanglement $C^{\prime}(t)=(C+\frac{1}{2})e^{-2\kappa
(t_{0}+t)}-\frac{1}{2}$. This can be simplified to $C^{\prime}(t)=\max\{0,
(C_{0}+\frac{1}{2})e^{-2\kappa t}-\frac{1}{2}\}$. It is obvious that this is
the same with the dynamics of entanglement for pure states with the same
initial entanglement $C_{0}$, as shown in Eq. (19). In fact, the familiar
Werner states belong to the decoherence path states.

Therefore in the depolarizing channels not only all pure states but also some
special mixed states have the same entanglement dynamics. In other words, the
pairwise entanglement of the decoherence path states are also the most stable.
We plot the dynamics of entanglement in decoherence path states with different
initial entanglement, as depicted in Fig. 2.

From Fig.2 (a) it can be seen that the entanglement in the two-qubit system
decreases with time due to its interaction with the decoherence environment.
There exists some time $T_{c}$ when its entanglement $C\left(  t\right)  =0$
for $t\geqslant T_{c}$. Thus $T_{c}$ is the critical time when the system
entanglement disappears, i.e. the two-qubit system becomes separable. We can
easily calculate the critical time:
\begin{equation}
T_{c}=\frac{\ln[ 2C_{0} +1]}{2\kappa}%
\end{equation}
The relation between $T_{c}$ and the initial entanglement $C_{0}$ is depicted
in Fig. 2 (b). Certainly, the lifetime of entanglement is longer if the
initial entanglement is larger. For the singlet state, the lifetime of
entanglement is $\kappa T_{c} =0.549.$

\begin{figure}[tbh]
\includegraphics[scale=0.5]{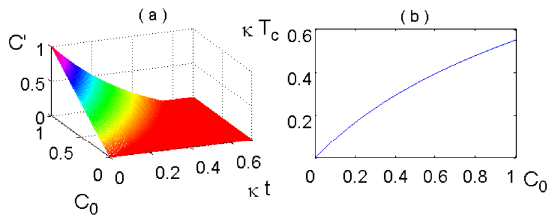}\caption{(Color online) The dynamics
of residual entanglement in a two-qubit system with initial decoherence path
states. (a) Residual entanglement $C^{\prime}$ as function of the initial
entanglement $C_{0}$ and time $t$. (b) Critical time $\kappa T_{c}$ as
function of the initial entanglement $C_{0}$. }%
\end{figure}

However, we can not exclude the possibility that there exist non-DPS states
with the same entanglement dynamics as DPS states. In the following, we can
see a concrete example. But as a special class of mixed states, the
decoherence path states of a given decoherence model are expected to exhibit
some other interesting properties.

\subsection{Ground State of Spin Models}

In this section we will investigate the stability of the nearest neighbor
entanglement of the ground states of spin models. The Hamiltonian of the
translationally invariant XXZ spin chain with periodic boundary condition is
\begin{equation}
H=\sum\nolimits_{i=1}^{N}[\sigma^{x}_{i}\sigma^{x}_{i+1}+\sigma^{y}_{i}%
\sigma^{y}_{i+1}+\gamma\sigma^{z}_{i}\sigma^{z}_{i+1}]
\end{equation}
when $\gamma=1$ the above Hamiltonian represents the Heisenberg
antiferromagnetic model. The ground state $\rho_{g}$\cite{Entangled
rings,Orbach} is translationally invariant and the total z component of spin
is zero. Thus the reduced density matrix of the $i$ and $i+1$ site is:
\begin{equation}
\rho_{i,i+1}=\left(
\begin{array}
[c]{cccc}%
u & 0 & 0 & 0\\
0 & x & z & 0\\
0 & z^{\ast} & y & 0\\
0 & 0 & 0 & v
\end{array}
\right)
\end{equation}
When $u=v$ and $x=y$ it corresponds to the ground state of the Heisenberg
antiferromagnetic model. The density matrices of the maximally entangled mixed
states \cite{Ishizaka,Moor,Munro} are also in the above form. In the general
Pauli channels of (6), the state $\rho_{i,i+1}$ evolve to
\begin{equation}
\rho^{\prime}_{i,i+1}=\left(
\begin{array}
[c]{cccc}%
A & 0 & 0 & E\\
0 & B & F & 0\\
0 & F^{\ast} & C & 0\\
E^{\ast} & 0 & 0 & D
\end{array}
\right)
\end{equation}
where $A=u\eta_{1}^{2}+v\eta_{2}^{2}+(x+y)\eta_{1}\eta_{2}$, $B=x\eta_{1}%
^{2}+y\eta_{2}^{2}+(u+v)\eta_{1}\eta_{2}$, $C=y\eta_{1}^{2}+x\eta_{2}%
^{2}+(u+v)\eta_{1}\eta_{2}$, $D=v\eta_{1}^{2}+u\eta_{2}^{2}+(x+y)\eta_{1}%
\eta_{2}$, $E=(z+z^{\ast})\eta_{3}\eta_{4}$ and $F=z\eta_{3}^{2}+z^{\ast}%
\eta_{4}^{2}$ with $\eta_{1}=p_{0}+p_{3}$, $\eta_{2}=p_{1}+p_{2}$, $\eta
_{3}=p_{0}-p_{3}$ and $\eta_{4}=p_{1}-p_{2}$. We note that $\eta_{1}\geq
|\eta_{3}|$ and $\eta_{2}\geq|\eta_{4}|$. In addition, $xy\geq|z|^{2}$ because
$\rho_{i,i+1}$ is positive. Thus $|E|\leq(BC)^{1/2}$. Therefore the
concurrence of $\rho^{\prime}_{i,i+1}$ is given by $C^{\prime}=2\max
\{0,|F|-(AD)^{1/2}\}$.

For the special Heisenberg antiferromagnetic model and the depolarizing
channels, the residual entanglement is $C^{\prime}=\max\{0, \eta_{3}^{2}%
C_{0}-2\eta_{1}\eta_{2}\}$, where $C_{0}=2(|z|-u)$. Note that $\eta
_{1}=(1+e^{-\kappa t})/2$, $\eta_{2}=(1-e^{-\kappa t})/2$ and $\eta
_{3}=e^{-\kappa t}$. Thus $C^{\prime}=\max\{0,(C_{0}+1/2)e^{-2\kappa t}-1/2
\}$, which is the same as the pure states. Therefore in the depolarizing
channels, the nearest neighbor entanglement of the ground state of the
Heisenberg antiferromagnetic model is the most stable, though it is not
maximized \cite{Entangled rings}. This result study the entanglement of the
ground states of spin models from a new point of view. In addition, we can
verify that $\rho_{i,i+1}$ does not belong to the decoherence path states.
This suggest that in this certain decoherence model, several mixed states
other than decoherence path states are also the most entanglement-stable.

\section{Conclusions and discussions}

In conclusion, we have investigate the entanglement dynamics of a two-qubit
system under a general decoherence model, that is Pauli channels. Given a
decoherence model, we find the special class of pure states that are the most
entanglement-stable and present the analytic entanglement dynamics of these
states. Since any pure states with the same entanglement are $LU$ equivalent,
we show that pure states are more favorable for maintaining entanglement than
general mixed states. Therefore in the situation of maintaining or
distributing entanglement, it is helpful to using pure entangled states and to
apply appropriate local unitary operations beforehand to transform the
entangled states to the most entanglement-stable form. Particularly, we
investigate a certain decoherence model i.e. the depolarizing channels. In
this case, a special class of mixed states that is decoherence path states are
as entanglement-stable as pure states. The familiar Werner states are indeed
DPS states. In addition, we investigate the entanglement dynamics of some
specific class of mixed states, such as the ground states of XXZ, Heisenberg
antiferromagnetic spin model and the maximally entangled mixed states. It is
shown that in the depolarizing channels, the nearest neighbor entanglement of
the ground state of Heisenberg antiferromagnetic spin model is coincidentally
the most entanglement-stable. Another interesting result is that if only one
qubit is coupled with the decoherence environment then the life time of
entanglement is independent on the initial entanglement. This just reflects
that entanglement is some kind of nonlocal property.

For the decoherence model considered in this paper, we find the most
entanglement-stable form of entangled states. The extension of this study to
more general decoherence models is very meaningful, which is also related to
the important work in \cite{Lidar2}. Furthermore, we also introduce a special
class mixed states i.e. decoherence path states and find that they are also
the most entanglement-stable states in the depolarizing channels. However,
they are expected to exhibit more interesting properties in general
decoherence models.

\begin{acknowledgments}
The authors thank Dr. Xiang-Fa Zhou for valuable discussions. This work was
funded by National Fundamental Research Program (2001CB309300), the Innovation
funds from Chinese Academy of Sciences, Z.-W. Zhou acknowledges funds from
National Natural Science Foundation of China (Grant No. 10204020).
\end{acknowledgments}

\end{document}